\documentclass[pre,aps,amsmath,amssymb,floatfix,twocolumn,longbibliography]{revtex4-1}

\pdfoutput=1
\usepackage{hyperref}
\usepackage{graphicx}
\usepackage[usenames]{color}
\usepackage{bm}

\definecolor{blue}{rgb}{0,0,0}
\newcommand{\bb}[1]{\textcolor{blue}{#1}}

\newcommand*\demon{\includegraphics[width=0.2cm]{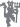}}

\def\cal#1{\mathcal{#1}}
\def\eqq#1{Eq.~(\ref{#1})}

\def\eq#1{(\ref{#1})}

\def\f#1{Fig.~\ref{#1}}

\def\c#1{~\cite{#1}}

\def\cc#1{Ref.~\cite{#1}}

\def\av#1{\langle #1 \rangle}

\def\beq{\begin{equation}}
\def\eeq{\end{equation}}
\def\bea{\begin{eqnarray}}
\def\eea{\end{eqnarray}}

\def\kt{k_{\rm B}T}
\def\tf{t_{\rm f}}

\begin{document}

\title{How to train your demon to do fast information erasure without heat production}
\author{Stephen Whitelam$^{\demon}$}\email{swhitelam@lbl.gov}
\affiliation{$^{\demon}$Molecular Foundry, Lawrence Berkeley National Laboratory, 1 Cyclotron Road, Berkeley, CA 94720, USA}

\begin{abstract}

Time-dependent protocols that perform irreversible logical operations, such as memory erasure, cost work and produce heat, placing bounds on the efficiency of computers. Here we use a prototypical computer model of a physical memory to show that it is possible to learn feedback-control protocols to do fast memory erasure without input of work or production of heat. These protocols, which are enacted by a neural-network ``demon'', do not violate the second law of thermodynamics because the demon generates more heat than the memory absorbs. The result is a form of nonlocal heat exchange in which one computation is rendered energetically favorable while a compensating one produces heat elsewhere, a tactic that could be used to rationally design the flow of energy within a computer.

\end{abstract}
\maketitle

\section{Introduction}

Landauer's principle states that erasing a bit of memory at temperature $T$ costs at least $\kt \ln 2$ units of work and generates the same amount of heat, placing bounds on the efficiency of computing\c{landauer1961irreversibility,piechocinska2000information,dillenschneider2009memory,jun2014high,berut2012experimental,dago2021information,hong2016experimental}. Moreover, these bounds are long-time limits: erasure protocols that occur in finite time $t_{\rm f}$ cost more energy than the Landauer bound by an amount that increases (to leading order) as $1/t_{\rm f}$\c{proesmans2020finite,zulkowski2014optimal,dago2022dynamics,konopik2023fundamental}. Thus, for a given physical system, the faster the computation the more heat it is likely to produce.

One way to locally suppress heat production during fast computation is to use fluctuating nanoscale elements controlled by feedback protocols. Such protocols can convert measurement information into stored heat and work\c{szilard1964decrease,sagawa2012nonequilibrium,esposito2012stochastic,parrondo2015thermodynamics,ehrich2022energetic,diana2013finite,toyabe2010experimental,abreu2011extracting}: Maxwell's demon, for instance, can create a heat engine by moving a partition in response to the fluctuations of gas molecules. Here we show that feedback protocols can exploit the thermal fluctuations of a model nanoscale memory unit to perform erasure without requiring work or producing heat, even if erasure happens far from equilibrium. Moreover, we show that such protocols can be learned in an autonomous, iterative way, using data accessible in a laboratory experiment. We consider a computer model of an overdamped colloidal particle in a double-well potential, a prototype of a single-bit memory. We use evolutionary methods to train a deep-neural-network demon that can alter the potential in response to input information. When the demon enacts a \bb{feedforward protocol, parameterized by time alone}, it learns effective and efficient erasure protocols that approach the Landauer bound as the time allotted for erasure is increased. When the demon also receives feedback from the system, the position of the colloidal particle, it can learn erasure protocols that extract work and store heat.

The result is a form of nonlocal heat exchange in which an irreversible computation can be performed in an energetically favorable way, with compensatory heat produced by the measurement apparatus and disposed of elsewhere. The learning algorithms used in this paper can be applied to experiment, and so the present results suggest a way of designing protocols to do computation with heat moved around a system for convenience. 

\section{Model and simulation details}

Consider a computer model of a colloidal particle in a potential, sketched in \f{fig0}. The particle has position $x$, and undergoes the Langevin dynamics
\beq
\label{langevin}
\dot{x}=-\partial_x U_{\bm c}(x) + \xi(t),
\eeq
where $\av{\xi(t)}=0$ and $\av{\xi(t) \xi(t')} = 2 \delta(t-t')$. Temperature is held constant throughout the paper, and we express energies in units of $\kt$. The potential $U_{\bm c}(x)$ is
\beq
\label{pot}
U_{\bm c}(x) = c_1 x+c_2 x^2+c_4 x^4,
\eeq
parameterized by the coefficients ${\bm c}=(c_1,c_2,c_4)$. This model defines a 1-bit memory, with its binary state $s=0,1$ determined by the location of the particle with respect to the origin, $s=\Theta(x)$~\footnote{We define $\Theta(x)=1$ if $x \geq 0$ and $\Theta(x) = 0$ otherwise.}.  At times $t=0$ and $t=\tf$ the potential parameters are ${\bm c}_0=(0,-2w,w)$, and so the potential has the double-well form shown in \f{fig0}(a). This has minima at $x=\pm 1$ and barrier height $w=5 \,\kt$. For times $0<t<\tf$, the parameters ${\bm c}$ are determined by a neural network. A neural network can represent an arbitrary function $U(x)$, if we allow full control of the system\c{proesmans2020optimal}, but here, to make closer contact with experiment, we consider the case of partial control, where only the coefficients ${\bm c}$ can be varied. In experiment, similar control can be effected by changing the power of a set of lasers\c{jun2014high}.
\begin{figure}[] 
   \centering
   \includegraphics[width=\linewidth]{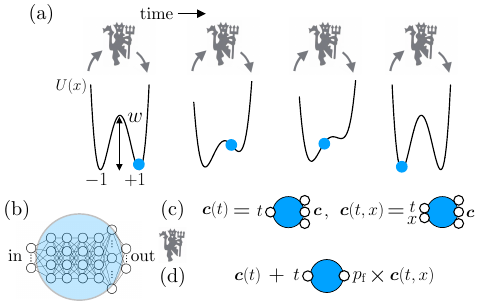} 
   \caption{Memory erasure by a neural-network demon. (a) We consider a particle undergoing Langevin dynamics in a potential $U_{\bm c}(x)$, \eqq{pot}, with the state of the memory given by $s=\Theta(x)$. A reset protocol is enacted by a neural-network demon, so that after time $\tf$ the system is in state $s=0$, regardless of its starting state. The demon controls the three coefficients ${\bm c}$ of the potential, similar to experiments that control the output power of a set of lasers. (b) All neural networks used in this paper have the same internal structure. (c) We consider demon protocols parameterized as a function of time; and time and position; and (d) a probabilistic combination of both.}
   \label{fig0}
\end{figure}

Dynamical trajectories are run for time $\tf$. We consider trajectories longer and shorter than the basic relaxation time $t_0=2$ of the system. The latter is the characteristic time for the particle, in the absence of a potential, to diffuse a distance equal to that between the potential well centers. (The relation of $t_0$ to experimental times depends on the details of the experiment: for example,~\cc{jun2014high} uses a 200 nm colloidal particle, whose diffusion constant is $D \sim (\mu {\rm m})^2/{\rm s}$, and a well-to-well separation of $2.5\, \mu$m, giving $t_0$ of order 10 seconds.)

At time $t=0$ the position $x_0$ of the particle is drawn from the Boltzmann distribution $\rho_0(x_0,{\bm c}_0)={\rm e}^{-U_{\bm c_0}(x_0)}/\int {\rm d}x' {\rm e}^{-U_{\bm c_0}(x')}$. Let $k = 1,2,\dots,K=\lfloor t_{\rm f}/\Delta t \rfloor$ label the simulation step. At regular time intervals $\Delta t=10^{-3}$ the potential coefficients are set to new values, ${\bm c}_k \to {\bm c}_{k+1}$, and the change of work recorded, $\Delta W_k= U_{{\bm c}_{k+1}}(x_k)-U_{{\bm c}_{k}}(x_k)$. The position of the particle is updated according to the forward Euler discretization of \eq{langevin}, 
\beq
x_{k+1}=x_k-\Delta t \, \partial_x U_{{\bm c}_{k+1}}(x) + \sqrt{2 \Delta t} \, \xi,
\eeq
where $\xi \sim {\cal N}(0,1)$, and the change of heat recorded, $\Delta Q_k=U_{{\bm c}_{k+1}}(x_{k+1})-U_{{\bm c}_{k+1}}(x_k)$. The total work and heat are $W=\sum_{k=0}^{K-1} \Delta W_k$ and $Q=\sum_{k=0}^{K-1} \Delta Q_k$. Note that ${\bm c}_K = {\bm c}_0$ and $W+Q = U_{{\bm c}_0}(x_K)-U_{{\bm c}_0}(x_0)$.

When the potential coefficients are updated, they are set to the values ${\bm c}= {\bm g}_{\bm \theta}({\bm i})$, determined by a deep neural network. Here ${\bm g}$ is the output 3-vector of the neural network, ${\bm i}$ is the input vector \bb{provided to the neural network}, and ${\bm \theta}$ is the vector of neural-network parameters (weights and biases). Details of the neural network are given in the attached code\c{erasure_github}. Briefly, it has an internal structure that is fully connected, with 4 hidden layers of width 4 and a final hidden layer of width 10, indicated by the blue circle in \f{fig0}(b). Neurons have tanh nonlinearities, and layer norm is used. This network structure is expressive enough to learn a range of functions, including rapidly-varying ones, and straightforward to train using non-gradient algorithms\c{whitelam2022training}. The arguments ${\bm i}$ are the inputs of the neural network. As sketched in \f{fig0}(c), we consider \bb{feedforward} protocols ${\bm c}(t)= (c_1(t), c_2(t), c_4(t))$ parameterized by time alone, in which case ${\bm i} = t/t_0$, and feedback protocols ${\bm c}(t,x)= (c_1(t,x), c_2(t,x), c_4(t,x))$ parameterized by time and particle position, in which case ${\bm i} = (t/t_0,x)$. The network has as many input neurons as input degrees of freedom. 

In an effort to learn protocols that achieve a particular objective, we use a genetic algorithm\c{GA,mitchell1998introduction,montana1989training,such2017deep} to adjust the parameters ${\bm \theta}$ of the neural network. This procedure is described in~\cc{whitelam2023demon}, and details are given in the attached code\c{erasure_github}. Briefly, we consider a population of 50 neural-network demons, each characterized by a value $\phi$ that is calculated by averaging a chosen quantity (see below) over $10^4$ independent trajectories. At the start of the evolutionary process, demon parameters are set equal to random numbers. The 5 demons with the smallest values of $\phi$ are chosen to be the parents of subsequent generations. These 5 are cloned and mutated, by adding random numbers to their parameters, to produce another population of 50. The procedure is repeated for several generations, the intent being to produce neural-network demons whose protocols give rise to values of $\phi$ as small as possible. For reversible protocols (particle translation and bit flipping) parameterized by time, this procedure recovers optimal-control results obtained analytically and by other numerical approaches\c{whitelam2023demon,schmiedl2007optimal,engel2022optimal}. In this paper we apply these methods to the irreversible operation of memory erasure. In \f{fig_s1} we benchmark the learning algorithm on a driven barrier-crossing problem closely related to the problem considered in the main text, showing that it reproduces the optimal-control protocols of~\cc{zhong2022limited}.

\section{Learning effective and efficient erasure protocols}

We first apply the evolutionary procedure to a \bb{feedforward protocol ${\bm c}(t)$, which is parameterized by time alone}. We choose to minimize the objective function
\beq
\label{phi1}
\phi=1-P_0+k_{\rm w} \av{W},
\eeq
where $P_0 \equiv \av{\Theta(-x(t_{\rm f}))}$ is the mean probability of reset. Here $\av{\cdot}$ denotes the mean over $10^4$ independent trajectories. Minimizing the term $1-P_0$ encourages protocols that reset the memory. Minimizing the final term in \eq{phi1} minimizes the mean work done under the protocol. The factor $k_{\rm w} = 0.05$ ensures that the final term in \eq{phi1} is smaller than $1-P_0$ until the probability of reset gets close to one, ensuring that the first goal of the evolutionary process is to produce a reset protocol, and the second goal is to do so with as little work as possible.

\f{fig1} shows the results of this procedure, for trajectory lengths $\tf=1,10$, and 100. As shown in panel (a), all result in protocols that achieve reset (with probability $\gtrsim 99.8 \%$) after sufficient evolutionary generations. These choices of $\tf$ correspond to a range from short to long, relative to the system's basic diffusive timescale $t_0=2$, and so present different challenges. Trajectories of length $\tf$ less than $t_0$ must involve strong driving, in order to move the particle the required distance in the allotted time. Gentler driving suffices for trajectories much longer than $t_0$, but effective reset protocols must still have a nonequilibrium character: the memory is volatile in the sense that the default energy barrier of $5 \, \kt$ can be readily surmounted by thermal fluctuations, and so energy input is required in order to maintain the fidelity of the memory. Assuming an escape time of $t_{\rm e} \approx \exp(5) \approx 150$, and modeling the potential as a 2-level system, memories that achieve reset and are then left in their final form for time $t_{\rm f}$ have a probability of remaining reset of approximately $(1+{\rm e}^{-2 t/t_{\rm e}})/2$, which is only $\approx 94\% $ or $\approx 63\% $ for $t_{\rm f}=10$ or $t_{\rm f}=100$, respectively. The learned protocols are considerably more effective.
\begin{figure}[] 
   \centering
   \includegraphics[width=\linewidth]{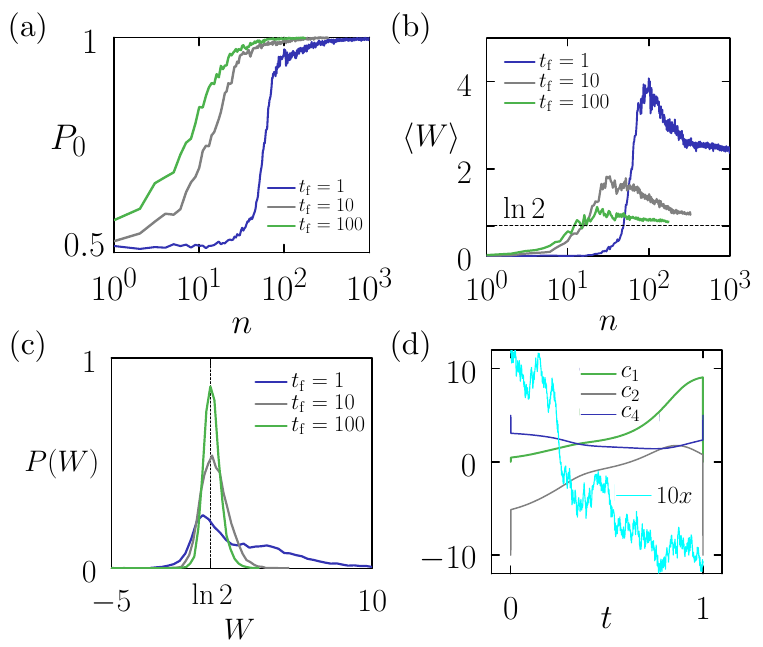} 
   \caption{Evolutionary learning of \bb{feedforward} erasure protocols parameterized by time, ${\bm c}(t)$. (a) Mean probability of erasure, $P_0$, for the best-performing protocol of generation $n$, for three different trajectory lengths $\tf$. Averages are taken over $10^4$ trajectories. (b) Mean work $\av{W}$ for the best-performing protocol of generation $n$. (c) Histograms $P(W)$ of work done for the best-performing protocol from the latest generation achieved for each $\tf$ (generations 1208, 327, and 175, for $\tf=$1, 10, and 100, respectively). In (b) and (c), the dotted line is the Landauer bound. (d) Erasure protocol ${\bm c}(t)$ for $\tf=1$ learned after 1208 evolutionary generations. The cyan line shows the trajectory of a single particle under this protocol.}
   \label{fig1}
\end{figure}

\f{fig1}(b) shows that the mean work $\av{W}$ required to achieve reset is close to the Landauer bound for $\tf=100$, and over $\kt$ away from the bound for $\tf=1$. These results are consistent with experiments done for times $t_{\rm f}$ larger and smaller than the system's basic relaxation time $t_0$\c{jun2014high}. Also consistent with those experiments, the work distribution resulting from the best protocol for each $\tf$ becomes increasingly sharply peaked about the Landauer bound as $\tf$ increases: see panel (c). Individual trajectories can achieve work values that violate the bound and even store work\c{jarzynski1997nonequilibrium,evans2002fluctuation,crooks1999entropy,seifert2012stochastic}. Panel (d) shows one of the learned time-dependent protocols, which is non-monotonic and displays jumps in the values of the potential coefficients at the start and end of the trajectory. Nonmonotonic features\c{zhong2022limited} and jumps\c{schmiedl2007optimal,blaber2021steps} are present in optimal-control protocols for other systems.

\begin{figure*}[] 
   \centering
   \includegraphics[width=\linewidth]{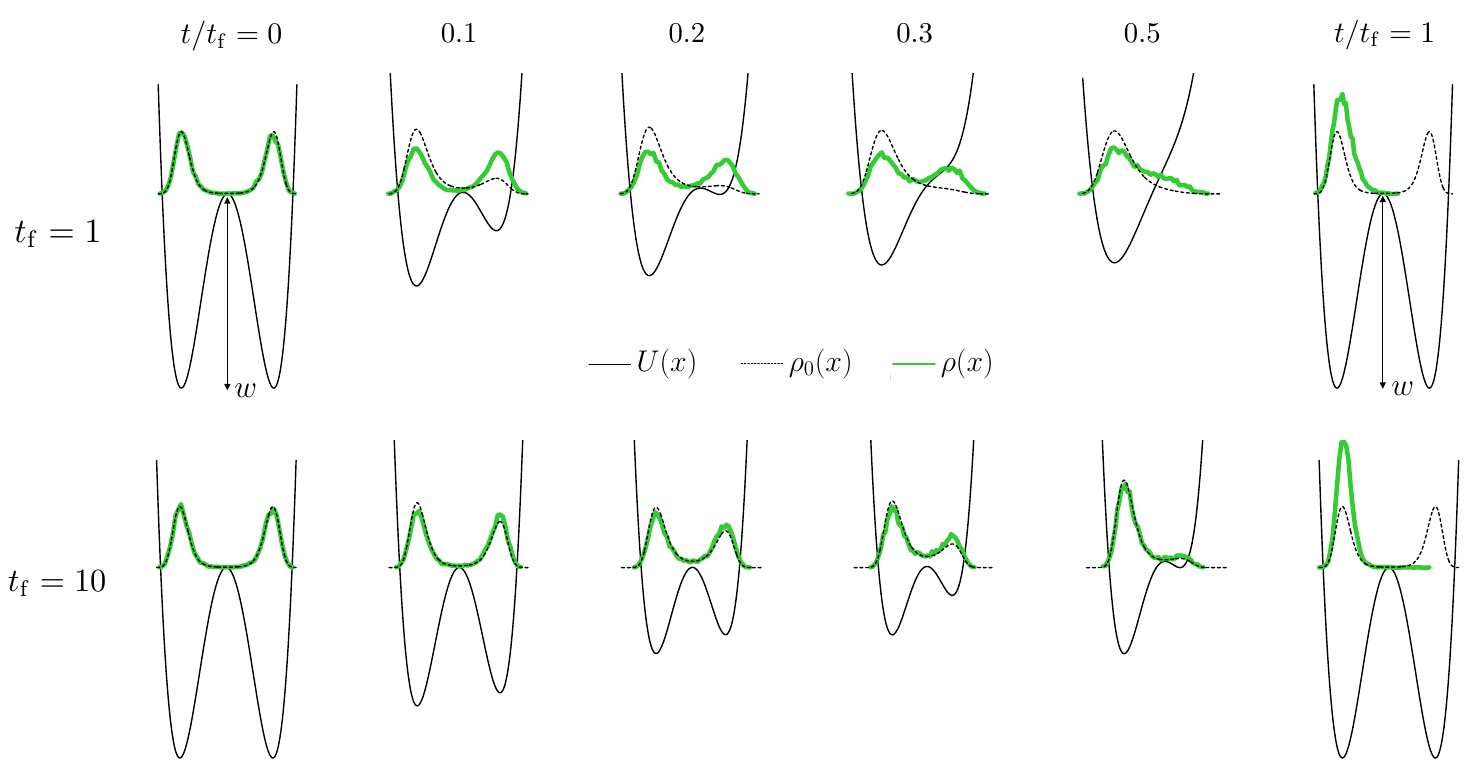} 
   \caption{Potentials $U(x)$ (black) and associated Boltzmann distributions $\rho_0(x)$ (dashed black) at 6 values of scaled time $t/\tf$, under feedforward protocols ${\bm c}(t)$ learned (after 1208 or 327 generations, respectively) by evolution for trajectory lengths $\tf=1$ (top) and $\tf=10$ (bottom). Histograms $\rho(x)$ of particle positions for $10^4$ independent trajectories are shown in green. The scale of the potential is set by the well depth $w=5 \kt$ at times $t=0, \tf$, and by the fact that the potential minima are at $x=\pm 1$. The common scale of the histograms and Boltzmann distributions is set by the Boltzmann distribution of $U(x)$ at times $t=0, \tf$, and by the fact that the area under each histogram is unity.}
   \label{fig4}
\end{figure*}

\f{fig4} illustrates the near-equilibrium and far-from-equilibrium character of the learned reset procedure for the cases $\tf=10$ and 1, respectively. For $\tf=10$, histograms of particle position $\rho(x)$ are, for most of the trajectory, close to the Boltzmann distribution $\rho_0(x,{\bm c})$ associated with the instantaneous potential $U_{\bm c}(x)$. At the end of the trajectory, these things become distinct: the particle-position histogram is consistent with a reset probability close to 1, while the Boltzmann distribution is consistent with a reset probability of $1/2$. For the case $\tf=1$, particle-position histograms and Boltzmann distributions are clearly different for most of the trajectory. Reset happens in a far-from-equilibrium way, requiring energy expenditure considerably in excess of the Landauer bound.

\section{Feedback protocols}

\bb{In this section we consider feedback protocols ${\bm c}(t,x)$ that are parameterized as a function of time $t$ and particle position $x$. Such protocols can be used to do fast erasure while violating the Landauer bound, if only the memory unit is considered. The cost of this violation is paid for by the acquisition (and subsequent erasure) of the particle coordinates, and so, in effect, one erasure process is used to control another erasure process. The advantage of doing so is that it is possible to render the chosen process energy efficient; the cost is that the controlling protocol is relatively inefficient. We shall also consider the question of how to learn probabilistic feedback-feedforward protocols that perform erasure with specified accuracy and efficiency, while making as few measurements as possible.}

 In \f{fig2}(a,b) we show that evolutionary learning applied to a demon fed both time $t$ and particle position $x$ can again produce effective protocols, with reset probabilities in excess of 99.9\%. Moreover, the mean work done upon reset is negative, with about $40 \, \kt$ extracted from the thermal bath in which the colloidal particle is immersed, even through the trajectory is short ($\tf=1$) relative to the basic diffusion time $t_0=2$. A similar amount of heat is absorbed from the bath.  \f{fig3} compares feedback protocols and feedforward (time-parameterized) protocols, showing that the former extract work from the bath through the small changes the demon makes to the potential coefficients in response to fluctuations of particle position. \bb{Panel (c) shows the behavior of the potential coefficients  for a single feedback-controlled trajectory in which the particle started in state $s=1$. \f{fig5} illustrates the shape of the potential as a function of time. (For the feedback protocol, the time evolution of the potential is different when the particle starts instead in state $s=0$: in that case, the potential minima fluctuate a little in order to extract work from the particle, but the potential shape does not change substantially.)}
 
\begin{figure}[] 
   \centering
   \includegraphics[width=\linewidth]{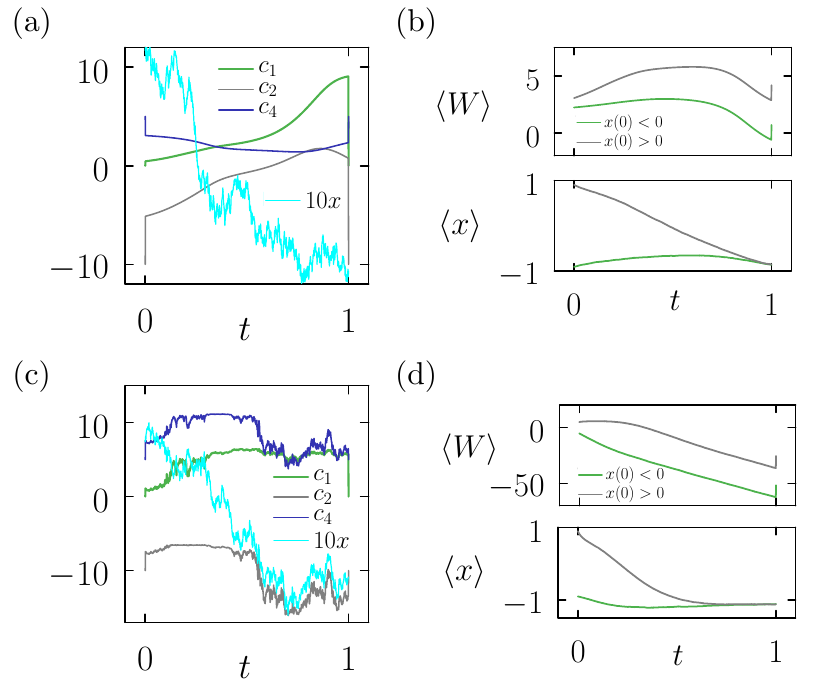} 
   \caption{Comparison of \bb{feedforward- (top) and feedback (bottom)} erasure protocols produced by evolutionary learning. Panel (a) is \f{fig1}(d), and panel (c) is its counterpart for a feedback protocol. Panels (b) and (d) show the mean work and mean position of particles under these protocols, separated according to whether the particle started in state $s=0$ (green) or $s=1$ (gray).}
   \label{fig3}
\end{figure}

\begin{figure*}[] 
   \centering
   \includegraphics[width=\linewidth]{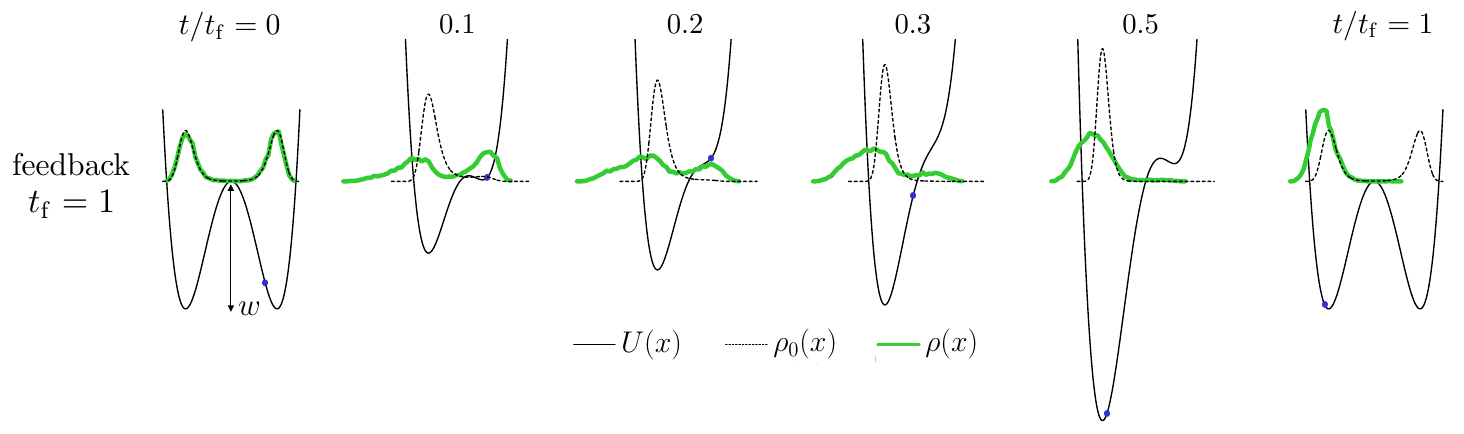} 
   \caption{Similar to \f{fig4}, but now for an evolution-learned feedback protocol ${\bm c}(t,x)$. With feedback, the potential $U(x)$ is different for each trajectory. We show the potential (black) \bb{from a trajectory in which the particle started in state $s=1$} (the instantaneous position of the particle is shown by a blue dot), and the associated Boltzmann distribution (black dashed) for that potential. Histograms $\rho(x)$ of particle positions for $10^4$ independent trajectories \bb{(starting with equal probability from either state)} are shown green.}
   \label{fig5}
\end{figure*}

Thus a feedback protocol can perform an irreversible logical operation in such a way that the surrounding thermal bath is cooled, rather than heated. To compensate, the demon must receive (and then erase) the real-valued colloidal coordinates it uses to perform the protocol. Assuming that each coordinate is represented as a 32-bit floating-point number, the demon requires at least $32 \kt \ln 2$ units of work to erase each measurement. For a trajectory of length $\tf=1$, the demon makes $10^3$ measurements, and so consumes at least $32,000 \, \kt \ln2$ units of work. With about $40 \, \kt$ stored in the memory, the efficiency of the system is then about $0.2\%$. The demon-memory system is therefore inefficient, but effective in the sense that the designated computation is done in an energetically favorable way, with waste heat sent elsewhere.

\section{Probabilistic feedback-feedforward protocols}

\bb{One way to reduce the information cost of a feedback protocol is to allow} the demon to choose when to make measurements. \bb{To do so, we can impose an evolutionary order parameter $\phi$ that encourages measurements to be made only when necessary, and allow the demon to choose, probabilistically, when to measure the particle coordinate $x$. This idea is sketched in \f{fig0}(d).} (The cost of measurement has also been considered in gradient-based approaches to reinforcement learning\c{bellinger2022balancing}, \bb{and the choice of stochastic versus deterministic control of a system has been explored in~\cc{kappen2005linear}}.) When the demon acts, it sets the potential coefficients $\bb{{\bm c}}$ to the values
\bea
\label{prob_protocol}
{\bm c} = \begin{cases}
  {\bm g}_{{\bm \theta}_1}(t)  & (\xi  \geq p_{\rm f}(t))\\
   {\bm g}_{{\bm \theta}_1}(t)+  {\bm g}_{{\bm \theta}_2}(t,x)  &  (\xi  < p_{\rm f}(t)).
\end{cases}
\eea
Here ${\bm g}_{{\bm \theta}_1}(t)$ and ${\bm g}_{{\bm \theta}_2}(t,x)$ are neural networks whose structures are shown in \f{fig0}(b,c)\bb{: each has three outputs, specifying the values of the potential coefficients ${\bm c}$, and a number of inputs equal to the number of input parameters it is fed (one for the case of the feedforward protocol ${\bm g}_{{\bm \theta}_1}(t)$, and two for the case of the feedback protocol ${\bm g}_{{\bm \theta}_2}(t,x)$).} $\xi$ is a uniform random real number on $(0,1]$, and $p_{\rm f}(t)=g_{{\bm \theta}_3}(t) \in [0,1]$ is the output of a third neural network \bb{(this has one input node, which accepts time $t$, and one output node, which specifies the probability $p_{\rm f}$).} Thus at time $t$ the demon decides with probability $p_{\rm f}$ to measure the particle position $x$ and update the potential coefficients using a feedback protocol. With probability $1-p_{\rm f}$ the demon makes no measurement, and instead updates ${\bm c}$ using a \bb{feedforward} protocol. 

\begin{figure}[] 
   \centering
   \includegraphics[width=\linewidth]{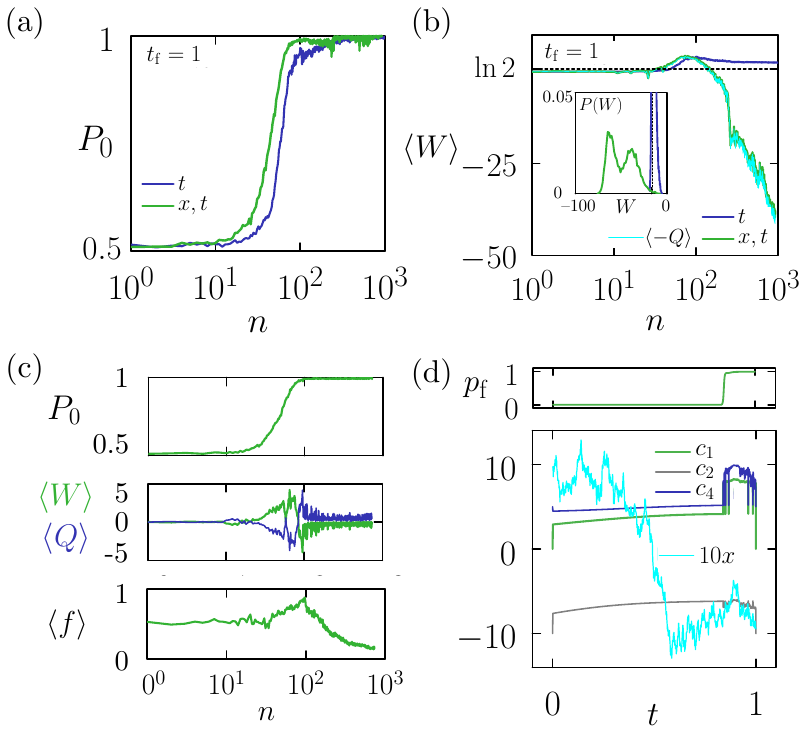} 
   \caption{Feedback erasure protocols produced by evolutionary learning. Panels (a) and (b) are similar to those of \f{fig1}, and include the time-dependent protocol for the case $\tf=1$ \bb{(blue, labeled ``$t$'')}, together with results obtained by training a feedback protocol ${\bm c}(t,x)$ \bb{(green, labeled ``$x,t$'').} The mean heat exchanged by the feedback protocol is shown in cyan: positive values of $Q$ indicate heat adsorption (negative entropy production). The inset to (b) shows the work distribution under the protocol learned after 922 generations (green), compared with that learned under the \bb{feedforward} protocol of \f{fig1} (blue). (c) Results obtained using a probabilistic feedback protocol \eq{prob_protocol} in which the demon can decide whether or not to make a measurement; $f$ is the number of measurements it makes, divided by the maximum possible number of measurements. (d) Trajectory produced using the best performing protocol from generation 722 of (c): $p_{\rm f}$ is the probability with which the demon makes a measurement.}
   \label{fig2}
\end{figure}

We instruct the learning algorithm to minimize the evolutionary order parameter
\bea
\label{phi2}
\phi= \begin{cases}
 \qquad \quad k_{\rm n} \av{n_{\rm f}}  &  \hspace{-1 cm} (P_0 > 0.99\, \, {\rm and} \, \, \av{W}< 0)\\
 1-P_0+k_{\rm w} \av{W}+k_{\rm c} & ({\rm otherwise}), 
\end{cases}
\eea
where $k_{\rm n}=3 \times 10^{-4}$, $k_{\rm w}=0.05$, and $k_{\rm c}=20$ are constants. Here $n_{\rm f}$ is the number of measurements of particle position made by the demon within a trajectory. \bb{The motivation behind~\eqq{phi2} is to imagine that we have a strict engineering requirement for a protocol with certain properties, such as to achieve reset with at least 99\% probability without energy expenditure, and then to ask for this to happen using as few measurements as possible (an alternative approach would be to consider the Pareto frontiers for combinations of these quantities).} The second clause in \eq{phi2} is \eq{phi1} shifted by a constant, and asks the demon to learn a reset protocol that requires as little work as possible. When it has produced a protocol with probability of reset $P_0>0.99$ that does not consume work, the first clause in \eq{phi2} becomes active and asks the demon to make as few measurements as possible. The constants $k$ ensure that the first clause in \eq{phi2} is always numerically smaller than the second \bb{-- as long as this is achieved, the precise numerical values of the constants are not important --} and so minimizing $\phi$ encourages the demon to maintain $P_0 \approx 0.99$ and $\av{W} \approx 0$ while making as few measurements as possible.

The results of this learning procedure are shown in \f{fig2}(c). As a function of evolutionary time $n$, the demon first learns to produce a reset protocol, and then to do so with no work consumed (and no heat expended, because $W \approx -Q$). During this stage of evolution it makes measurements more frequently in order to reduce the work done during reset. Subsequently, it learns to maintain a reset protocol with $P_0 \approx 0.99$ and $\av{W} \approx 0$ while making fewer measurements. In this phase it learns to make a fraction $f \approx 15\%$ of the measurements it would have made in a pure feedback protocol. \f{fig2}(d) shows that its strategy for doing so is strongly time-dependent. For the majority of the trajectory the demon maintains the measurement-feedback probability $p_{\rm f}$ close to zero, and enacts a \bb{feedforward} protocol. Close to the end of the trajectory it sets $p_{\rm f}$ close to one, and makes frequent measurement-feedback decisions in order to make the reset process energetically neutral. 

\section{Conclusions}

We have shown that evolutionary methods can train a deep-neural-network demon to identify reset protocols for a noisy and volatile memory element, on timescales longer and shorter than its basic relaxation time. These protocols are effective, achieving reset with probability close to one, and are energy-efficient, approaching the Landauer bound as protocol time is increased. For times short compared to the basic relaxation time of the system, reset requires work considerably in excess of the Landauer bound, consistent with experiments. In this short-time regime, we have shown that feedback protocols can effect reset without heat production. In this mode the demon performs nonlocal heat exchange, generating heat and doing work by making measurements and deleting data, while the memory element under its control produces no heat or stores it. 

The evolutionary learning method used to train the demon is technically simple and effective. It reproduces optimal-control results obtained by other methods, for \bb{feedforward} protocols parameterized by time alone, and can identify feedback protocols that satisfy multiple objectives of fidelity and efficiency. Given that it takes as input time- and ensemble averages of fluctuating quantities that are accessible experimentally, the present results suggest a way of designing experimental protocols that do computation with heat moved around a system for convenience.

\section{Acknowledgments}

I thank Adrianne Zhong for providing the data from~\cc{zhong2022limited} labeled ``OCT'' in \f{fig_s1}, and thank Corneel Casert and Isaac Tamblyn for discussions. Code for doing evolutionary learning of a feedback erasure protocol can be found at~\cc{erasure_github}. This work was performed at the Molecular Foundry at Lawrence Berkeley National Laboratory, supported by the Office of Basic Energy Sciences of the U.S. Department of Energy under Contract No. DE-AC02--05CH11231.

\appendix
\section{Benchmarking the evolutionary learning algorithm}

\setcounter{figure}{0}
\renewcommand{\thefigure}{A\arabic{figure}}

In \f{fig_s1} we benchmark the evolutionary learning algorithm on the driven barrier-crossing problem of~\cc{zhong2022limited}; see panels (b) and (d) of Fig. 2 of that paper. A particle undergoes the Langevin dynamics \eq{langevin}, with $U_{\bm c}(x)$ replaced by the the potential $U_\lambda(x) = 16 [(x^2-1)^2/4-\lambda x]$, in units of $\kt$. The initial and final values of the control parameter $\bb{\lambda}$ are $\lambda(0)=-1$ and $\lambda(t_{\rm f})=1$, and we train a neural network to express $\lambda(t)$ for $0<t<t_{\rm f}$ so as to minimize $\phi=\av{W}$. Each particle begins in equilibrium in the left-hand potential well, and averages are calculated over $10^4$ independent trajectories. Our results are consistent with the optimal-control solutions of~\cc{zhong2022limited}.

\begin{figure}[h] 
   \centering
   \includegraphics[width=\linewidth]{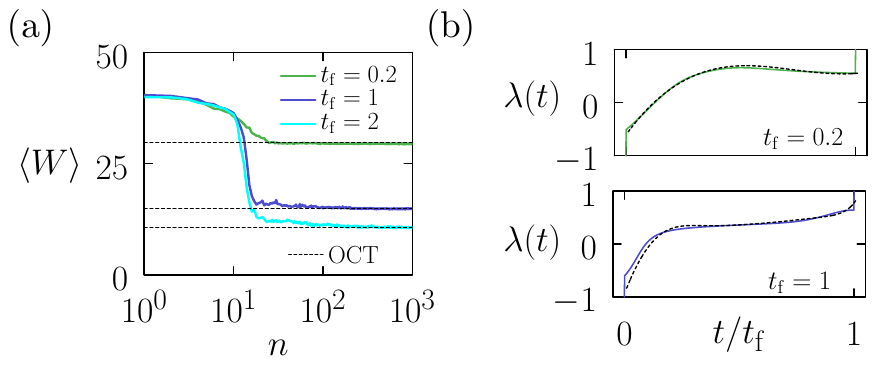} 
   \caption{Benchmarking the learning algorithm using the driven barrier-crossing problem of~\cc{zhong2022limited}. (a) $\av{W}$ for the best-performing protocol as a function of evolutionary time $n$, for three values of $t_{\rm f}$ (colors); these converge to the values calculated by the optimal control theory (OCT) approach of~\cc{zhong2022limited} (dashed black). (b) Learned protocols as a function of time for two values of $t_{\rm f}$. These converge to forms that fluctuate (with increasing generation) about the noise-free OCT solutions (dashed black). For $t_{\rm f}=0.2$ the profile is nonmonotonic, consistent with the results of~\cc{zhong2022limited}. }
   \label{fig_s1}
\end{figure}


\begin{thebibliography}{38}%
\makeatletter
\providecommand \@ifxundefined [1]{%
 \@ifx{#1\undefined}
}%
\providecommand \@ifnum [1]{%
 \ifnum #1\expandafter \@firstoftwo
 \else \expandafter \@secondoftwo
 \fi
}%
\providecommand \@ifx [1]{%
 \ifx #1\expandafter \@firstoftwo
 \else \expandafter \@secondoftwo
 \fi
}%
\providecommand \natexlab [1]{#1}%
\providecommand \enquote  [1]{``#1''}%
\providecommand \bibnamefont  [1]{#1}%
\providecommand \bibfnamefont [1]{#1}%
\providecommand \citenamefont [1]{#1}%
\providecommand \href@noop [0]{\@secondoftwo}%
\providecommand \href [0]{\begingroup \@sanitize@url \@href}%
\providecommand \@href[1]{\@@startlink{#1}\@@href}%
\providecommand \@@href[1]{\endgroup#1\@@endlink}%
\providecommand \@sanitize@url [0]{\catcode `\\12\catcode `\$12\catcode
  `\&12\catcode `\#12\catcode `\^12\catcode `\_12\catcode `\%12\relax}%
\providecommand \@@startlink[1]{}%
\providecommand \@@endlink[0]{}%
\providecommand \url  [0]{\begingroup\@sanitize@url \@url }%
\providecommand \@url [1]{\endgroup\@href {#1}{\urlprefix }}%
\providecommand \urlprefix  [0]{URL }%
\providecommand \Eprint [0]{\href }%
\providecommand \doibase [0]{http://dx.doi.org/}%
\providecommand \selectlanguage [0]{\@gobble}%
\providecommand \bibinfo  [0]{\@secondoftwo}%
\providecommand \bibfield  [0]{\@secondoftwo}%
\providecommand \translation [1]{[#1]}%
\providecommand \BibitemOpen [0]{}%
\providecommand \bibitemStop [0]{}%
\providecommand \bibitemNoStop [0]{.\EOS\space}%
\providecommand \EOS [0]{\spacefactor3000\relax}%
\providecommand \BibitemShut  [1]{\csname bibitem#1\endcsname}%
\let\auto@bib@innerbib\@empty
\bibitem [{\citenamefont {Landauer}(1961)}]{landauer1961irreversibility}%
  \BibitemOpen
  \bibfield  {author} {\bibinfo {author} {\bibfnamefont {Rolf}\ \bibnamefont
  {Landauer}},\ }\bibfield  {title} {\enquote {\bibinfo {title}
  {Irreversibility and heat generation in the computing process},}\ }\href@noop
  {} {\bibfield  {journal} {\bibinfo  {journal} {IBM journal of research and
  development}\ }\textbf {\bibinfo {volume} {5}},\ \bibinfo {pages} {183--191}
  (\bibinfo {year} {1961})}\BibitemShut {NoStop}%
\bibitem [{\citenamefont {Piechocinska}(2000)}]{piechocinska2000information}%
  \BibitemOpen
  \bibfield  {author} {\bibinfo {author} {\bibfnamefont {Barbara}\ \bibnamefont
  {Piechocinska}},\ }\bibfield  {title} {\enquote {\bibinfo {title}
  {Information erasure},}\ }\href@noop {} {\bibfield  {journal} {\bibinfo
  {journal} {Physical Review A}\ }\textbf {\bibinfo {volume} {61}},\ \bibinfo
  {pages} {062314} (\bibinfo {year} {2000})}\BibitemShut {NoStop}%
\bibitem [{\citenamefont {Dillenschneider}\ and\ \citenamefont
  {Lutz}(2009)}]{dillenschneider2009memory}%
  \BibitemOpen
  \bibfield  {author} {\bibinfo {author} {\bibfnamefont {Raoul}\ \bibnamefont
  {Dillenschneider}}\ and\ \bibinfo {author} {\bibfnamefont {Eric}\
  \bibnamefont {Lutz}},\ }\bibfield  {title} {\enquote {\bibinfo {title}
  {Memory erasure in small systems},}\ }\href@noop {} {\bibfield  {journal}
  {\bibinfo  {journal} {Physical Review Letters}\ }\textbf {\bibinfo {volume}
  {102}},\ \bibinfo {pages} {210601} (\bibinfo {year} {2009})}\BibitemShut
  {NoStop}%
\bibitem [{\citenamefont {Jun}\ \emph {et~al.}(2014)\citenamefont {Jun},
  \citenamefont {Gavrilov},\ and\ \citenamefont {Bechhoefer}}]{jun2014high}%
  \BibitemOpen
  \bibfield  {author} {\bibinfo {author} {\bibfnamefont {Yonggun}\ \bibnamefont
  {Jun}}, \bibinfo {author} {\bibfnamefont {Mom{\v{c}}ilo}\ \bibnamefont
  {Gavrilov}}, \ and\ \bibinfo {author} {\bibfnamefont {John}\ \bibnamefont
  {Bechhoefer}},\ }\bibfield  {title} {\enquote {\bibinfo {title}
  {High-precision test of {L}andauer{'}s principle in a feedback trap},}\
  }\href@noop {} {\bibfield  {journal} {\bibinfo  {journal} {Physical Review
  Letters}\ }\textbf {\bibinfo {volume} {113}},\ \bibinfo {pages} {190601}
  (\bibinfo {year} {2014})}\BibitemShut {NoStop}%
\bibitem [{\citenamefont {B{\'e}rut}\ \emph {et~al.}(2012)\citenamefont
  {B{\'e}rut}, \citenamefont {Arakelyan}, \citenamefont {Petrosyan},
  \citenamefont {Ciliberto}, \citenamefont {Dillenschneider},\ and\
  \citenamefont {Lutz}}]{berut2012experimental}%
  \BibitemOpen
  \bibfield  {author} {\bibinfo {author} {\bibfnamefont {Antoine}\ \bibnamefont
  {B{\'e}rut}}, \bibinfo {author} {\bibfnamefont {Artak}\ \bibnamefont
  {Arakelyan}}, \bibinfo {author} {\bibfnamefont {Artyom}\ \bibnamefont
  {Petrosyan}}, \bibinfo {author} {\bibfnamefont {Sergio}\ \bibnamefont
  {Ciliberto}}, \bibinfo {author} {\bibfnamefont {Raoul}\ \bibnamefont
  {Dillenschneider}}, \ and\ \bibinfo {author} {\bibfnamefont {Eric}\
  \bibnamefont {Lutz}},\ }\bibfield  {title} {\enquote {\bibinfo {title}
  {Experimental verification of {L}andauer{'}s principle linking information
  and thermodynamics},}\ }\href@noop {} {\bibfield  {journal} {\bibinfo
  {journal} {Nature}\ }\textbf {\bibinfo {volume} {483}},\ \bibinfo {pages}
  {187--189} (\bibinfo {year} {2012})}\BibitemShut {NoStop}%
\bibitem [{\citenamefont {Dago}\ \emph {et~al.}(2021)\citenamefont {Dago},
  \citenamefont {Pereda}, \citenamefont {Barros}, \citenamefont {Ciliberto},\
  and\ \citenamefont {Bellon}}]{dago2021information}%
  \BibitemOpen
  \bibfield  {author} {\bibinfo {author} {\bibfnamefont {Salamb{\^o}}\
  \bibnamefont {Dago}}, \bibinfo {author} {\bibfnamefont {Jorge}\ \bibnamefont
  {Pereda}}, \bibinfo {author} {\bibfnamefont {Nicolas}\ \bibnamefont
  {Barros}}, \bibinfo {author} {\bibfnamefont {Sergio}\ \bibnamefont
  {Ciliberto}}, \ and\ \bibinfo {author} {\bibfnamefont {Ludovic}\ \bibnamefont
  {Bellon}},\ }\bibfield  {title} {\enquote {\bibinfo {title} {Information and
  thermodynamics: fast and precise approach to {L}andauer{'}s bound in an
  underdamped micromechanical oscillator},}\ }\href@noop {} {\bibfield
  {journal} {\bibinfo  {journal} {Physical Review Letters}\ }\textbf {\bibinfo
  {volume} {126}},\ \bibinfo {pages} {170601} (\bibinfo {year}
  {2021})}\BibitemShut {NoStop}%
\bibitem [{\citenamefont {Hong}\ \emph {et~al.}(2016)\citenamefont {Hong},
  \citenamefont {Lambson}, \citenamefont {Dhuey},\ and\ \citenamefont
  {Bokor}}]{hong2016experimental}%
  \BibitemOpen
  \bibfield  {author} {\bibinfo {author} {\bibfnamefont {Jeongmin}\
  \bibnamefont {Hong}}, \bibinfo {author} {\bibfnamefont {Brian}\ \bibnamefont
  {Lambson}}, \bibinfo {author} {\bibfnamefont {Scott}\ \bibnamefont {Dhuey}},
  \ and\ \bibinfo {author} {\bibfnamefont {Jeffrey}\ \bibnamefont {Bokor}},\
  }\bibfield  {title} {\enquote {\bibinfo {title} {Experimental test of
  {L}andauer{'}s principle in single-bit operations on nanomagnetic memory
  bits},}\ }\href@noop {} {\bibfield  {journal} {\bibinfo  {journal} {Science
  advances}\ }\textbf {\bibinfo {volume} {2}},\ \bibinfo {pages} {e1501492}
  (\bibinfo {year} {2016})}\BibitemShut {NoStop}%
\bibitem [{\citenamefont {Proesmans}\ \emph
  {et~al.}(2020{\natexlab{a}})\citenamefont {Proesmans}, \citenamefont
  {Ehrich},\ and\ \citenamefont {Bechhoefer}}]{proesmans2020finite}%
  \BibitemOpen
  \bibfield  {author} {\bibinfo {author} {\bibfnamefont {Karel}\ \bibnamefont
  {Proesmans}}, \bibinfo {author} {\bibfnamefont {Jannik}\ \bibnamefont
  {Ehrich}}, \ and\ \bibinfo {author} {\bibfnamefont {John}\ \bibnamefont
  {Bechhoefer}},\ }\bibfield  {title} {\enquote {\bibinfo {title} {Finite-time
  {L}andauer principle},}\ }\href@noop {} {\bibfield  {journal} {\bibinfo
  {journal} {Physical Review Letters}\ }\textbf {\bibinfo {volume} {125}},\
  \bibinfo {pages} {100602} (\bibinfo {year} {2020}{\natexlab{a}})}\BibitemShut
  {NoStop}%
\bibitem [{\citenamefont {Zulkowski}\ and\ \citenamefont
  {DeWeese}(2014)}]{zulkowski2014optimal}%
  \BibitemOpen
  \bibfield  {author} {\bibinfo {author} {\bibfnamefont {Patrick~R}\
  \bibnamefont {Zulkowski}}\ and\ \bibinfo {author} {\bibfnamefont {Michael~R}\
  \bibnamefont {DeWeese}},\ }\bibfield  {title} {\enquote {\bibinfo {title}
  {Optimal finite-time erasure of a classical bit},}\ }\href@noop {} {\bibfield
   {journal} {\bibinfo  {journal} {Physical Review E}\ }\textbf {\bibinfo
  {volume} {89}},\ \bibinfo {pages} {052140} (\bibinfo {year}
  {2014})}\BibitemShut {NoStop}%
\bibitem [{\citenamefont {Dago}\ and\ \citenamefont
  {Bellon}(2022)}]{dago2022dynamics}%
  \BibitemOpen
  \bibfield  {author} {\bibinfo {author} {\bibfnamefont {Salamb{\^o}}\
  \bibnamefont {Dago}}\ and\ \bibinfo {author} {\bibfnamefont {Ludovic}\
  \bibnamefont {Bellon}},\ }\bibfield  {title} {\enquote {\bibinfo {title}
  {Dynamics of information erasure and extension of {L}andauer{'}s bound to
  fast processes},}\ }\href@noop {} {\bibfield  {journal} {\bibinfo  {journal}
  {Physical Review Letters}\ }\textbf {\bibinfo {volume} {128}},\ \bibinfo
  {pages} {070604} (\bibinfo {year} {2022})}\BibitemShut {NoStop}%
\bibitem [{\citenamefont {Konopik}\ \emph {et~al.}(2023)\citenamefont
  {Konopik}, \citenamefont {Korten}, \citenamefont {Lutz},\ and\ \citenamefont
  {Linke}}]{konopik2023fundamental}%
  \BibitemOpen
  \bibfield  {author} {\bibinfo {author} {\bibfnamefont {Michael}\ \bibnamefont
  {Konopik}}, \bibinfo {author} {\bibfnamefont {Till}\ \bibnamefont {Korten}},
  \bibinfo {author} {\bibfnamefont {Eric}\ \bibnamefont {Lutz}}, \ and\
  \bibinfo {author} {\bibfnamefont {Heiner}\ \bibnamefont {Linke}},\ }\bibfield
   {title} {\enquote {\bibinfo {title} {Fundamental energy cost of finite-time
  parallelizable computing},}\ }\href@noop {} {\bibfield  {journal} {\bibinfo
  {journal} {Nature Communications}\ }\textbf {\bibinfo {volume} {14}},\
  \bibinfo {pages} {447} (\bibinfo {year} {2023})}\BibitemShut {NoStop}%
\bibitem [{\citenamefont {Szilard}(1964)}]{szilard1964decrease}%
  \BibitemOpen
  \bibfield  {author} {\bibinfo {author} {\bibfnamefont {Leo}\ \bibnamefont
  {Szilard}},\ }\bibfield  {title} {\enquote {\bibinfo {title} {On the decrease
  of entropy in a thermodynamic system by the intervention of intelligent
  beings},}\ }\href@noop {} {\bibfield  {journal} {\bibinfo  {journal}
  {Behavioral Science}\ }\textbf {\bibinfo {volume} {9}},\ \bibinfo {pages}
  {301--310} (\bibinfo {year} {1964})}\BibitemShut {NoStop}%
\bibitem [{\citenamefont {Sagawa}\ and\ \citenamefont
  {Ueda}(2012)}]{sagawa2012nonequilibrium}%
  \BibitemOpen
  \bibfield  {author} {\bibinfo {author} {\bibfnamefont {Takahiro}\
  \bibnamefont {Sagawa}}\ and\ \bibinfo {author} {\bibfnamefont {Masahito}\
  \bibnamefont {Ueda}},\ }\bibfield  {title} {\enquote {\bibinfo {title}
  {Nonequilibrium thermodynamics of feedback control},}\ }\href@noop {}
  {\bibfield  {journal} {\bibinfo  {journal} {Physical Review E}\ }\textbf
  {\bibinfo {volume} {85}},\ \bibinfo {pages} {021104} (\bibinfo {year}
  {2012})}\BibitemShut {NoStop}%
\bibitem [{\citenamefont {Esposito}\ and\ \citenamefont
  {Schaller}(2012)}]{esposito2012stochastic}%
  \BibitemOpen
  \bibfield  {author} {\bibinfo {author} {\bibfnamefont {Massimiliano}\
  \bibnamefont {Esposito}}\ and\ \bibinfo {author} {\bibfnamefont {Gernot}\
  \bibnamefont {Schaller}},\ }\bibfield  {title} {\enquote {\bibinfo {title}
  {Stochastic thermodynamics for ``{M}axwell demon'' feedbacks},}\ }\href@noop
  {} {\bibfield  {journal} {\bibinfo  {journal} {EPL (Europhysics Letters)}\
  }\textbf {\bibinfo {volume} {99}},\ \bibinfo {pages} {30003} (\bibinfo {year}
  {2012})}\BibitemShut {NoStop}%
\bibitem [{\citenamefont {Parrondo}\ \emph {et~al.}(2015)\citenamefont
  {Parrondo}, \citenamefont {Horowitz},\ and\ \citenamefont
  {Sagawa}}]{parrondo2015thermodynamics}%
  \BibitemOpen
  \bibfield  {author} {\bibinfo {author} {\bibfnamefont {Juan~MR}\ \bibnamefont
  {Parrondo}}, \bibinfo {author} {\bibfnamefont {Jordan~M}\ \bibnamefont
  {Horowitz}}, \ and\ \bibinfo {author} {\bibfnamefont {Takahiro}\ \bibnamefont
  {Sagawa}},\ }\bibfield  {title} {\enquote {\bibinfo {title} {Thermodynamics
  of information},}\ }\href@noop {} {\bibfield  {journal} {\bibinfo  {journal}
  {Nature physics}\ }\textbf {\bibinfo {volume} {11}},\ \bibinfo {pages}
  {131--139} (\bibinfo {year} {2015})}\BibitemShut {NoStop}%
\bibitem [{\citenamefont {Ehrich}\ \emph {et~al.}(2022)\citenamefont {Ehrich},
  \citenamefont {Still},\ and\ \citenamefont {Sivak}}]{ehrich2022energetic}%
  \BibitemOpen
  \bibfield  {author} {\bibinfo {author} {\bibfnamefont {Jannik}\ \bibnamefont
  {Ehrich}}, \bibinfo {author} {\bibfnamefont {Susanne}\ \bibnamefont {Still}},
  \ and\ \bibinfo {author} {\bibfnamefont {David~A}\ \bibnamefont {Sivak}},\
  }\bibfield  {title} {\enquote {\bibinfo {title} {Energetic cost of feedback
  control},}\ }\href@noop {} {\bibfield  {journal} {\bibinfo  {journal} {arXiv
  preprint arXiv:2206.10793}\ } (\bibinfo {year} {2022})}\BibitemShut {NoStop}%
\bibitem [{\citenamefont {Diana}\ \emph {et~al.}(2013)\citenamefont {Diana},
  \citenamefont {Bagci},\ and\ \citenamefont {Esposito}}]{diana2013finite}%
  \BibitemOpen
  \bibfield  {author} {\bibinfo {author} {\bibfnamefont {Giovanni}\
  \bibnamefont {Diana}}, \bibinfo {author} {\bibfnamefont {G~Baris}\
  \bibnamefont {Bagci}}, \ and\ \bibinfo {author} {\bibfnamefont
  {Massimiliano}\ \bibnamefont {Esposito}},\ }\bibfield  {title} {\enquote
  {\bibinfo {title} {Finite-time erasing of information stored in fermionic
  bits},}\ }\href@noop {} {\bibfield  {journal} {\bibinfo  {journal} {Physical
  Review E}\ }\textbf {\bibinfo {volume} {87}},\ \bibinfo {pages} {012111}
  (\bibinfo {year} {2013})}\BibitemShut {NoStop}%
\bibitem [{\citenamefont {Toyabe}\ \emph {et~al.}(2010)\citenamefont {Toyabe},
  \citenamefont {Sagawa}, \citenamefont {Ueda}, \citenamefont {Muneyuki},\ and\
  \citenamefont {Sano}}]{toyabe2010experimental}%
  \BibitemOpen
  \bibfield  {author} {\bibinfo {author} {\bibfnamefont {Shoichi}\ \bibnamefont
  {Toyabe}}, \bibinfo {author} {\bibfnamefont {Takahiro}\ \bibnamefont
  {Sagawa}}, \bibinfo {author} {\bibfnamefont {Masahito}\ \bibnamefont {Ueda}},
  \bibinfo {author} {\bibfnamefont {Eiro}\ \bibnamefont {Muneyuki}}, \ and\
  \bibinfo {author} {\bibfnamefont {Masaki}\ \bibnamefont {Sano}},\ }\bibfield
  {title} {\enquote {\bibinfo {title} {Experimental demonstration of
  information-to-energy conversion and validation of the generalized jarzynski
  equality},}\ }\href@noop {} {\bibfield  {journal} {\bibinfo  {journal}
  {Nature physics}\ }\textbf {\bibinfo {volume} {6}},\ \bibinfo {pages}
  {988--992} (\bibinfo {year} {2010})}\BibitemShut {NoStop}%
\bibitem [{\citenamefont {Abreu}\ and\ \citenamefont
  {Seifert}(2011)}]{abreu2011extracting}%
  \BibitemOpen
  \bibfield  {author} {\bibinfo {author} {\bibfnamefont {David}\ \bibnamefont
  {Abreu}}\ and\ \bibinfo {author} {\bibfnamefont {Udo}\ \bibnamefont
  {Seifert}},\ }\bibfield  {title} {\enquote {\bibinfo {title} {Extracting work
  from a single heat bath through feedback},}\ }\href@noop {} {\bibfield
  {journal} {\bibinfo  {journal} {Europhysics Letters}\ }\textbf {\bibinfo
  {volume} {94}},\ \bibinfo {pages} {10001} (\bibinfo {year}
  {2011})}\BibitemShut {NoStop}%
\bibitem [{Note1()}]{Note1}%
  \BibitemOpen
  \bibinfo {note} {We define $\Theta (x)=1$ if $x \geq 0$ and $\Theta (x) = 0$
  otherwise.}\BibitemShut {Stop}%
\bibitem [{\citenamefont {Proesmans}\ \emph
  {et~al.}(2020{\natexlab{b}})\citenamefont {Proesmans}, \citenamefont
  {Ehrich},\ and\ \citenamefont {Bechhoefer}}]{proesmans2020optimal}%
  \BibitemOpen
  \bibfield  {author} {\bibinfo {author} {\bibfnamefont {Karel}\ \bibnamefont
  {Proesmans}}, \bibinfo {author} {\bibfnamefont {Jannik}\ \bibnamefont
  {Ehrich}}, \ and\ \bibinfo {author} {\bibfnamefont {John}\ \bibnamefont
  {Bechhoefer}},\ }\bibfield  {title} {\enquote {\bibinfo {title} {Optimal
  finite-time bit erasure under full control},}\ }\href@noop {} {\bibfield
  {journal} {\bibinfo  {journal} {Physical Review E}\ }\textbf {\bibinfo
  {volume} {102}},\ \bibinfo {pages} {032105} (\bibinfo {year}
  {2020}{\natexlab{b}})}\BibitemShut {NoStop}%
\bibitem [{era()}]{erasure_github}%
  \BibitemOpen
  \href@noop {} {}\bibinfo {howpublished}
  {\url{https://github.com/swhitelam/erasure}}\BibitemShut {NoStop}%
\bibitem [{\citenamefont {Whitelam}\ \emph {et~al.}(2022)\citenamefont
  {Whitelam}, \citenamefont {Selin}, \citenamefont {Benlolo}, \citenamefont
  {Casert},\ and\ \citenamefont {Tamblyn}}]{whitelam2022training}%
  \BibitemOpen
  \bibfield  {author} {\bibinfo {author} {\bibfnamefont {Stephen}\ \bibnamefont
  {Whitelam}}, \bibinfo {author} {\bibfnamefont {Viktor}\ \bibnamefont
  {Selin}}, \bibinfo {author} {\bibfnamefont {Ian}\ \bibnamefont {Benlolo}},
  \bibinfo {author} {\bibfnamefont {Corneel}\ \bibnamefont {Casert}}, \ and\
  \bibinfo {author} {\bibfnamefont {Isaac}\ \bibnamefont {Tamblyn}},\
  }\bibfield  {title} {\enquote {\bibinfo {title} {Training neural networks
  using {M}etropolis {M}onte {C}arlo and an adaptive variant},}\ }\href@noop {}
  {\bibfield  {journal} {\bibinfo  {journal} {Machine Learning: Science and
  Technology}\ }\textbf {\bibinfo {volume} {3}},\ \bibinfo {pages} {045026}
  (\bibinfo {year} {2022})}\BibitemShut {NoStop}%
\bibitem [{\citenamefont {Holland}(1992)}]{GA}%
  \BibitemOpen
  \bibfield  {author} {\bibinfo {author} {\bibfnamefont {John~H}\ \bibnamefont
  {Holland}},\ }\bibfield  {title} {\enquote {\bibinfo {title} {Genetic
  algorithms},}\ }\href@noop {} {\bibfield  {journal} {\bibinfo  {journal}
  {Scientific American}\ }\textbf {\bibinfo {volume} {267}},\ \bibinfo {pages}
  {66--73} (\bibinfo {year} {1992})}\BibitemShut {NoStop}%
\bibitem [{\citenamefont {Mitchell}(1998)}]{mitchell1998introduction}%
  \BibitemOpen
  \bibfield  {author} {\bibinfo {author} {\bibfnamefont {Melanie}\ \bibnamefont
  {Mitchell}},\ }\href@noop {} {\emph {\bibinfo {title} {An introduction to
  genetic algorithms}}}\ (\bibinfo  {publisher} {MIT press},\ \bibinfo {year}
  {1998})\BibitemShut {NoStop}%
\bibitem [{\citenamefont {Montana}\ and\ \citenamefont
  {Davis}(1989)}]{montana1989training}%
  \BibitemOpen
  \bibfield  {author} {\bibinfo {author} {\bibfnamefont {David~J}\ \bibnamefont
  {Montana}}\ and\ \bibinfo {author} {\bibfnamefont {Lawrence}\ \bibnamefont
  {Davis}},\ }\bibfield  {title} {\enquote {\bibinfo {title} {Training
  feedforward neural networks using genetic algorithms.}}\ }in\ \href@noop {}
  {\emph {\bibinfo {booktitle} {IJCAI}}},\ Vol.~\bibinfo {volume} {89}\
  (\bibinfo {year} {1989})\ pp.\ \bibinfo {pages} {762--767}\BibitemShut
  {NoStop}%
\bibitem [{\citenamefont {Such}\ \emph {et~al.}(2017)\citenamefont {Such},
  \citenamefont {Madhavan}, \citenamefont {Conti}, \citenamefont {Lehman},
  \citenamefont {Stanley},\ and\ \citenamefont {Clune}}]{such2017deep}%
  \BibitemOpen
  \bibfield  {author} {\bibinfo {author} {\bibfnamefont {Felipe~Petroski}\
  \bibnamefont {Such}}, \bibinfo {author} {\bibfnamefont {Vashisht}\
  \bibnamefont {Madhavan}}, \bibinfo {author} {\bibfnamefont {Edoardo}\
  \bibnamefont {Conti}}, \bibinfo {author} {\bibfnamefont {Joel}\ \bibnamefont
  {Lehman}}, \bibinfo {author} {\bibfnamefont {Kenneth~O}\ \bibnamefont
  {Stanley}}, \ and\ \bibinfo {author} {\bibfnamefont {Jeff}\ \bibnamefont
  {Clune}},\ }\bibfield  {title} {\enquote {\bibinfo {title} {Deep
  neuroevolution: genetic algorithms are a competitive alternative for training
  deep neural networks for reinforcement learning},}\ }\href@noop {} {\bibfield
   {journal} {\bibinfo  {journal} {arXiv preprint arXiv:1712.06567}\ }
  (\bibinfo {year} {2017})}\BibitemShut {NoStop}%
\bibitem [{\citenamefont {Whitelam}(2023)}]{whitelam2023demon}%
  \BibitemOpen
  \bibfield  {author} {\bibinfo {author} {\bibfnamefont {Stephen}\ \bibnamefont
  {Whitelam}},\ }\bibfield  {title} {\enquote {\bibinfo {title} {Demon in the
  machine: learning to extract work and absorb entropy from fluctuating
  nanosystems},}\ }\href@noop {} {\bibfield  {journal} {\bibinfo  {journal}
  {Physical Review X}\ }\textbf {\bibinfo {volume} {13}},\ \bibinfo {pages}
  {021005} (\bibinfo {year} {2023})}\BibitemShut {NoStop}%
\bibitem [{\citenamefont {Schmiedl}\ and\ \citenamefont
  {Seifert}(2007)}]{schmiedl2007optimal}%
  \BibitemOpen
  \bibfield  {author} {\bibinfo {author} {\bibfnamefont {Tim}\ \bibnamefont
  {Schmiedl}}\ and\ \bibinfo {author} {\bibfnamefont {Udo}\ \bibnamefont
  {Seifert}},\ }\bibfield  {title} {\enquote {\bibinfo {title} {Optimal
  finite-time processes in stochastic thermodynamics},}\ }\href@noop {}
  {\bibfield  {journal} {\bibinfo  {journal} {Physical Review Letters}\
  }\textbf {\bibinfo {volume} {98}},\ \bibinfo {pages} {108301} (\bibinfo
  {year} {2007})}\BibitemShut {NoStop}%
\bibitem [{\citenamefont {Engel}\ \emph {et~al.}(2022)\citenamefont {Engel},
  \citenamefont {Smith},\ and\ \citenamefont {Brenner}}]{engel2022optimal}%
  \BibitemOpen
  \bibfield  {author} {\bibinfo {author} {\bibfnamefont {Megan~C}\ \bibnamefont
  {Engel}}, \bibinfo {author} {\bibfnamefont {Jamie~A}\ \bibnamefont {Smith}},
  \ and\ \bibinfo {author} {\bibfnamefont {Michael~P}\ \bibnamefont
  {Brenner}},\ }\bibfield  {title} {\enquote {\bibinfo {title} {Optimal control
  of nonequilibrium systems through automatic differentiation},}\ }\href@noop
  {} {\bibfield  {journal} {\bibinfo  {journal} {arXiv preprint
  arXiv:2201.00098}\ } (\bibinfo {year} {2022})}\BibitemShut {NoStop}%
\bibitem [{\citenamefont {Zhong}\ and\ \citenamefont
  {DeWeese}(2022)}]{zhong2022limited}%
  \BibitemOpen
  \bibfield  {author} {\bibinfo {author} {\bibfnamefont {Adrianne}\
  \bibnamefont {Zhong}}\ and\ \bibinfo {author} {\bibfnamefont {Michael~R}\
  \bibnamefont {DeWeese}},\ }\bibfield  {title} {\enquote {\bibinfo {title}
  {Limited-control optimal protocols arbitrarily far from equilibrium},}\
  }\href@noop {} {\bibfield  {journal} {\bibinfo  {journal} {Physical Review
  E}\ }\textbf {\bibinfo {volume} {106}},\ \bibinfo {pages} {044135} (\bibinfo
  {year} {2022})}\BibitemShut {NoStop}%
\bibitem [{\citenamefont {Jarzynski}(1997)}]{jarzynski1997nonequilibrium}%
  \BibitemOpen
  \bibfield  {author} {\bibinfo {author} {\bibfnamefont {Christopher}\
  \bibnamefont {Jarzynski}},\ }\bibfield  {title} {\enquote {\bibinfo {title}
  {Nonequilibrium equality for free energy differences},}\ }\href@noop {}
  {\bibfield  {journal} {\bibinfo  {journal} {Physical Review Letters}\
  }\textbf {\bibinfo {volume} {78}},\ \bibinfo {pages} {2690} (\bibinfo {year}
  {1997})}\BibitemShut {NoStop}%
\bibitem [{\citenamefont {Evans}\ and\ \citenamefont
  {Searles}(2002)}]{evans2002fluctuation}%
  \BibitemOpen
  \bibfield  {author} {\bibinfo {author} {\bibfnamefont {Denis~J}\ \bibnamefont
  {Evans}}\ and\ \bibinfo {author} {\bibfnamefont {Debra~J}\ \bibnamefont
  {Searles}},\ }\bibfield  {title} {\enquote {\bibinfo {title} {The fluctuation
  theorem},}\ }\href@noop {} {\bibfield  {journal} {\bibinfo  {journal}
  {Advances in Physics}\ }\textbf {\bibinfo {volume} {51}},\ \bibinfo {pages}
  {1529--1585} (\bibinfo {year} {2002})}\BibitemShut {NoStop}%
\bibitem [{\citenamefont {Crooks}(1999)}]{crooks1999entropy}%
  \BibitemOpen
  \bibfield  {author} {\bibinfo {author} {\bibfnamefont {Gavin~E}\ \bibnamefont
  {Crooks}},\ }\bibfield  {title} {\enquote {\bibinfo {title} {Entropy
  production fluctuation theorem and the nonequilibrium work relation for free
  energy differences},}\ }\href@noop {} {\bibfield  {journal} {\bibinfo
  {journal} {Physical Review E}\ }\textbf {\bibinfo {volume} {60}},\ \bibinfo
  {pages} {2721} (\bibinfo {year} {1999})}\BibitemShut {NoStop}%
\bibitem [{\citenamefont {Seifert}(2012)}]{seifert2012stochastic}%
  \BibitemOpen
  \bibfield  {author} {\bibinfo {author} {\bibfnamefont {Udo}\ \bibnamefont
  {Seifert}},\ }\bibfield  {title} {\enquote {\bibinfo {title} {Stochastic
  thermodynamics, fluctuation theorems and molecular machines},}\ }\href@noop
  {} {\bibfield  {journal} {\bibinfo  {journal} {Reports on Progress in
  Physics}\ }\textbf {\bibinfo {volume} {75}},\ \bibinfo {pages} {126001}
  (\bibinfo {year} {2012})}\BibitemShut {NoStop}%
\bibitem [{\citenamefont {Blaber}\ \emph {et~al.}(2021)\citenamefont {Blaber},
  \citenamefont {Louwerse},\ and\ \citenamefont {Sivak}}]{blaber2021steps}%
  \BibitemOpen
  \bibfield  {author} {\bibinfo {author} {\bibfnamefont {Steven}\ \bibnamefont
  {Blaber}}, \bibinfo {author} {\bibfnamefont {Miranda~D}\ \bibnamefont
  {Louwerse}}, \ and\ \bibinfo {author} {\bibfnamefont {David~A}\ \bibnamefont
  {Sivak}},\ }\bibfield  {title} {\enquote {\bibinfo {title} {Steps minimize
  dissipation in rapidly driven stochastic systems},}\ }\href@noop {}
  {\bibfield  {journal} {\bibinfo  {journal} {Physical Review E}\ }\textbf
  {\bibinfo {volume} {104}},\ \bibinfo {pages} {L022101} (\bibinfo {year}
  {2021})}\BibitemShut {NoStop}%
\bibitem [{\citenamefont {Bellinger}\ \emph {et~al.}(2022)\citenamefont
  {Bellinger}, \citenamefont {Drozdyuk}, \citenamefont {Crowley},\ and\
  \citenamefont {Tamblyn}}]{bellinger2022balancing}%
  \BibitemOpen
  \bibfield  {author} {\bibinfo {author} {\bibfnamefont {Colin}\ \bibnamefont
  {Bellinger}}, \bibinfo {author} {\bibfnamefont {Andriy}\ \bibnamefont
  {Drozdyuk}}, \bibinfo {author} {\bibfnamefont {Mark}\ \bibnamefont
  {Crowley}}, \ and\ \bibinfo {author} {\bibfnamefont {Isaac}\ \bibnamefont
  {Tamblyn}},\ }\bibfield  {title} {\enquote {\bibinfo {title} {Balancing
  information with observation costs in deep reinforcement learning},}\ }in\
  \href@noop {} {\emph {\bibinfo {booktitle} {Proceedings of the 35th Canadian
  Conference on Artificial Intelligence. CAIAC}}}\ (\bibinfo {year}
  {2022})\BibitemShut {NoStop}%
\bibitem [{\citenamefont {Kappen}(2005)}]{kappen2005linear}%
  \BibitemOpen
  \bibfield  {author} {\bibinfo {author} {\bibfnamefont {Hilbert~J}\
  \bibnamefont {Kappen}},\ }\bibfield  {title} {\enquote {\bibinfo {title}
  {Linear theory for control of nonlinear stochastic systems},}\ }\href@noop {}
  {\bibfield  {journal} {\bibinfo  {journal} {Physical Review Letters}\
  }\textbf {\bibinfo {volume} {95}},\ \bibinfo {pages} {200201} (\bibinfo
  {year} {2005})}\BibitemShut {NoStop}%
\end{thebibliography}

%

\end{document}